\documentstyle[12pt]{article}
\begin{document}
\baselineskip=24pt
\begin{flushright}
Preprint No: IMSc/98/01/03\\
\end{flushright}
\begin{center}
{\bf A HUNDRED YEARS OF LARMOR FORMULA }

\vskip .5cm

{\bf K.H.MARIWALLA AND N.D.HARI DASS} \\
{\it Institute of Mathematical Sciences} \\
{\it CIT Campus, CHENNAI - 600 113, INDIA}
\end{center} 

\vskip 1cm

\centerline{\bf ABSTRACT}

\vskip .5cm

Sir Joseph LARMOR showed in 1897 that an oscillating electric charge emits radiation energy
proportional to (acceleration)$^2$. At first sight,the result appears to
be valid for arbitrary accelerations.
But, perpetual uniform acceleration has been a case of nagging doubts, as radiation
reaction vanishes and the equivalence principle, as also conformal symmetry of Maxwell
equations each require nil energy loss. Special hypotheses are
devised by some to justify the assumption of radiation loss for both perpetual and 
non-perpetual (uniform) accelerations which, as in the case of (uniform) velocities,
are really different. The problem is here simply resolved by an explicit computation to show absence of
radiation for the perpetual case and by illustrating that Larmor formula
makes sense {\it only if} there is {\it change} in acceleration, just as
kinetic energy has nontrivial quantitative sense, only when there is change in velocity.

\newpage

Many a breakthrough came to be announced in 1897. 
Besides J.J.Thompson's initial discovery of the electron and Planck's reluctant
recognition of an essential role of Boltzman's viewpoint leading him in 1899 to
his radiation law, Sir Joseph Larmor had two papers of particular historical note. 
One$^1$, third in the series `{\it On Dynamical Theory of the 
Electric and Luminiferous Medium}', introduced
the notions of time-dilation and length contraction with complete transformations of
space-time (now known after Lorentz) 
and of electromagnetic fields to appear in his Adams Prize Essay$^2$. The other$^3$
proposed an explanation of Zeeman effect (discovered in 1896) different from
Lorentz, giving his famous precision effect on a rotating charge in a magnetic
field. The same paper also gives the celebrated Larmor formula for the rate at 
which an oscillator of charge $e$ emits radiation energy, viz.
$$
{\cal R} \ = \ \frac{2}{3} \frac{e^2}{c^3} {\bf g}^2 \ ; \eqno(1)
$$
${\bf g} = {\bf \dot{v}}$ is the acceleration and $c$ the velocity of light, which
at times, apparent from the context, will be taken as 1. Whereas
the subject of the first paper above was completely clarified in Einstein's
Special Relativity (SR), and the Larmor precision effect subsumed in quantum theory,
the significance of Larmor formula (1) continues to be a
subject of debate. For instance, retarded field$^{4,5}$ of a charge, in arbitrary
acceleration, gives the above radiation rate for both relativistic and
non-relativistic cases.  Lorentz showed that energy loss in non-relativistic case is 
equivalent to a force of `damping' or `radiation reaction' :
$$
{\bf \Gamma} \ = \ \left(\frac{2}{3} \frac{e^2}{c^3}\right) {\bf \ddot{v}} \ ; \eqno(2)
$$
As ${\bf\ddot{v}} = 0$ for perpetual uniform acceleration (PUA), one
would conclude that there is no radiation
in this case. However radiation should occur if uniform acceleration is for a finite
period T, as ${\bf \ddot{v}} \ne 0$ at the end points. Two other arguments for 
absence of radiation  for PUA are suggested by the equivalance principle (EP) and
conformal symmetry of the motion.

\vskip .5cm

\noindent (A) \ Weak and Strong versions of Equivalence Principle$^{6,7,8,9}$ are: (WEP)
Path of a structureless test body is a geodesic ; (SEP) local
space-time geometry is of Special Relativity (SR). These hold in the
large for uniform (= static, homogeneous) gravitational field (UGF) over extended
regions, for the following idealised situations. WEP-A uniformly accelerated frame 
of acceleration -{\bf g} is
equivalent to one supported in an UGF of field intensity {\bf g} ; SEP - A freely
falling frame in an UGF is equivalent to an inertial frame in SR. In particular,
electric charges placed in the above accelerated or freely falling frames would not
be seen to radiate, as ones supported in static gravitational fields or inertial
frames are understood not to radiate. In the above, equivalent
means indistinguishable {\it only} as a regards experiments within the specified frames. To
verify the conclusion in other frames one must admit as basic the
reality of radiation and its dual role as {\it the}
fundamental means of communication with particulate structure in accord with SR and quantum theory. 
e.g. 1) far-field of a charge once disengaged , exists as a
free-field or photons independent of the source,and capable of
'darkening a photo-plate' irreversibly ;
trapped between two reflecting, parallel filters A,B it constitutes a 
macroscopic clock in constant relation with a clock,say, at A. 2)
State of motion of an observer cannot trigger or suppress emission of radiation by a
distant source.3) Radiation may not leave a source without leaving
footprints, as it carries away both linear and angular momenta. 
4) Fields can not transmit through space-time if the only fields present
are static. Then, according to EP a charge in PUA does not radiate in all
generality.

\vskip .5cm

\noindent (B) \ Relativistic generalization of (2) was given by 
Abraham as a 4-vector orthogonal to 4-velocity v$^\nu$ and 4-acceleration 
a$^\mu$:
$$
\Gamma^u \ = \ \frac{2}{3} \frac{e^2}{c^3} G^\mu \ , \ G^\mu \ =
\ \frac{da^\mu}{d\tau} \ -a^2 v^\mu \eqno(3)
$$
where $a^2 = a^\mu a_\mu = {\bf g}^2 > 0$ and
$\tau$ is proper time. This expression is also standard in the works
of P.A.M. Dirac and of C.J.Eliezer. As $G^\mu = 0$ is the condition
for uniform acceleration, there is no radiation reaction or
radiation loss. Now, both $G^\mu = 0$ and Maxwell free field
equations have for their invariance group the 15 parameter
conformal group$^{10,11,12}$. Of the four special conformal
transformations, three with space-like parameters {\it hike
accelerations} just as three pure Lorentz transformations {\it boost
velocities}, while the fourth , with time-like parameter, like space-time
dilations, induces electromagnetic gauge transformations.
Thus the wavezone or radiation field is unaffected by
acceleration hikes, so a charge in uniform acceleration can not
radiate as one at rest does not. It is amusing that if one
replaces, in the Poincare group, Lorentz boosts and time
translations by acceleration hikes and space-time dilations,   one
obtains$^{12}$ the ten parameter (de Sitter) group transitive on the
complete connected homogeneous space-time with the metric $(c^2
dt^2 - d{\bf x}^2) t^{-2}$, instead of $c^2dt^2 - d{\bf x}^2$ for
the Poincare group. This space-time is thus a model of an UGF,
unique upto isomorphism. In particular one can analyse observer paths along, and on
neighbouring geodesics to study applications to EP. But this is well outside the
scope of this work.

\vskip .5cm

An overwhelming majority of scientists, led by G.A.Schott$^{13}$,
Fritz Rohrlich$^{14,15}$, Sydney Coleman ({\it Classical Electron Theory From a
Modern Standpoint}, Rand. Corp. preprint RM-2820-PR Sept. 1961) and Rudolf
Peierls$^{16}$ have argued that since $g \ne 0$ in the Larmor
formula, there is radiation and therefore a paradox to be
resolved. For this, varied new hypotheses are introduced in
the nature of ignotum per ignotus : 
1) {\it Whether acceleration is perpetual or for a short period is irrelevant$^{14,15}$,
as the first is only asymptotic}. Such an argument for velocities 
also results in a
paradox : the clock paradox of SR.
2) {\it Energy assumed lost comes from the time-rate of change of a specially postulated 
"Acceleration$^{13}$ or Schott$^{14}$ energy",
$\frac{2}{3} \frac{e^2}{c^3}  a_0$, of a charge}. It forces$^{9}$ radiation reaction 
in (3) to be the timelike vector ${\cal R}v^\mu$.
3) {\it The relevance of the equivalence principle to this issue is 
controversial$^{15}$} ; `the question whether
a charge really radiates is meaningless'$^{15}$, and depends on observer acceleration,
vanishing in co-accelerating frames$^{15}$. Such an extreme standpoint is not even
in accord with classical and quantum measurement concepts. 
4) Conformal acceleration hikes are ill-suited and replaced$^{15}$ by
coordinate change $dT^2 - dZ^2 - dx^2 - dy^2 = z^2 dt^2 - (dz/g)^2 - dx^2
- dy^2$ used by Pauli$^{17}$ to obtain field of an accelerated charge. Rindler coordinate
horizons of this metric are invoked$^{16}$ to render radiation
unobservable. The analogous coordinate change $dT^2-dZ^2-dx^2-dy^2=(dt/u)^2-t^2 dz^2
- dx^2 -dy^2$ substituting Lorentz boosts, begs interpretation$^{9,18}$. 
Some recent studies of `absorption criteria' and `particle detectors' for radiation,
both classical and quantum, for accelerated frames show absence of field
excitations or thermal background$^{19,20,21,22}$.

\vskip .5cm

Standpoint of the present work, in accord with Milner$^{23}$, Pauli$^{17}$, 
von Laue$^{24}$, Bondi$^{25,26}$, Schwinger (Lecture 6, Sec.34 in the manuscript
{\it Classical Electrodynamics} by J.Schwinger, L.L.De Raad Jr., K.A.Milton, W.Y.Tsai
- 1979), and others$^{8,9,11,12,18}$, is that the 
matter is not at all recondite :
the proponents of radiation hypothesis are making an {\it
unobtrusive} assumption that Larmor formula is a necessary and
sufficient condition for radiation, which it is not. In fact, 
as shown here, radiation occurs only if there is change in
acceleration, so much so that if uniform acceleration is for a
period $T$ only, the frequency spectrum of radiation is centred
at frequency $\omega = 2\pi/T$. Hence, for perpetual uniform
acceleration there is no radiation loss at all. Quite independent of such 
arguments, an exact calculation using 
the complete (Bondi-Gold)$^{26,27}$ fields of a charge in PUA give zero radiation 
rate$^{8,9,25}$. The viewpoint given here was first exposited$^{8,27}$ at a meeting
of the American Physical Society in March 1965 with considerable detail in {\it
Classical Radiation From A Uniformly Accelerated Charge}, preprints (with JJ
Walker-referred to as MW) University of Wisconsin (Madison, 1965), and Indian Institute
of Technology (Bombay, 1966) ; reference (9) gives a short exposition. Coleman's
work, agreeing essentially with Rohrlich's, came to attention only recently. The
powerful use, here, of the technique of `Spectral resolution' is inspired by
Schwinger,  who uses it with effect for the non-relativistic case.

\vskip .5cm

Consider a charged particle of velocity $v_o$, which is given, at
time $t=0$, uniform acceleration {\bf g} for a period $T$ ; then
$$
{\bf v}(t) = {\bf v}_o + {\bf g}t \Theta (t) - {\bf g} (t-T)
\Theta (t-T)\,. \eqno(4)
$$
$$
{\bf \dot{ v}}(t) = {\bf g} (\Theta(t) - \Theta (t-T)),
\ddot{\bf v}(t) = {\bf g} (\delta(t) - \delta(t-T)) \eqno(5)
$$
where $\Theta, \delta$ are Heaviside-unit and Dirac-delta
functions. Modulo the rate at which charges do work on the fields $\int {\bf
E} \cdot {\bf J} dV$ and the identity
$$
\frac{\partial U}{\partial t} + {\bf E} \cdot {\bf J} \ = \ -
\bigtriangledown \cdot {\bf S}\,, \eqno(6)
$$
the rate of damping energy
$$
{\bf v} . \Gamma \ = \ \frac{d}{dt} ({\bf v} \cdot \dot{\bf v}) -
{\cal R}
$$
equals the time rate of field energy $W = \int U dV, U = ({\bf E}^2 + {\bf
H}^2)/8\pi$, in a volume surrounding the charge, or of the outgoing flux $P = \int {\bf
S}. d {\bf \Sigma}$, accross the bounding surface, where ${\bf S} = c {\bf E} \times {\bf H}/4\pi$
is the Poynting vector. Clearly ${\bf v} \cdot \ddot{\bf v}$ is not well defined due to
distribution products ; taking $x\delta(x) = 0$ one perceives end
point effects, which become explicit on intergration over all $t$.
Discarding the term integrated out and using Plancheral theorem
(integral of the absolute square of a function equals the integral of the
the absolute square of
its Fourier transform).
$$
\int {\cal R}(t) dt \ = \ \frac{2}{3} \frac{c^2}{c^3}
\int^{+\infty}_{-\infty} \dot{v}^2 dt = \frac{2}{3}
\frac{e^2}{c^3} \int^{+\infty}_{-\infty} (\hat{\dot{v}}(\omega))^2 \ =
\ \int \hat{\cal R}(\omega) d\omega \eqno(7)
$$
where $\hat{v}(\omega) =  \int^{+\infty}_{-\infty} e^{-i\omega t} v(t) dt /
\sqrt{2\pi}, \hat{\dot{v}}(\omega) = -i\omega\hat{v}(\omega)$, and 
$$
\hat{\ddot{v}}(\omega) \ = \ -i\omega\dot{v}(\omega) \ = \ \frac{-2ig}{\sqrt{2\pi}}
\sin \frac{\omega T}{2} e^{i\omega T/2} \eqno(8)
$$
giving the spectral distribution of outgoing power radiated
$$
\hat{\cal R}(\omega) \ = \ \frac{1}{\pi} \cdot \frac{2}{3} \frac{e^2}{c^3}
{\bf g}^2 \left( \frac{\sin \omega T/2}{\omega/2}\right)^2 \,. \eqno(9)
$$
This has maxima at $\omega = 0, \frac{3\pi}{T}, \frac{5\pi}{T}$  ..... with decreasing 
amplitude, showing interference of contributions at two end points : more
apart the end points, smaller the frequency $\omega_0\stackrel{def} {=} 2\pi/T$. In the
limit, as $T \rightarrow \infty$, the (power) spectrum is seen to
be concentrated at $\omega=0$, so that there is no radiation at all in
PUA. This simple picture does not quite survive in the relativistic domain, though the 
conclusion of no radiation in PUA holds in all generality.

\vskip .5cm

\noindent {\bf The Complete Fields} 

\vskip .5cm

With the Lorentz condition imposed (for details, upto eqn.(17) see MW), 
the integral expressions for the retarded
4-potentials$^{4}$ for an arbitrarily moving charge `e' at $Q$ are
$$
A_\mu (x) \ = \ 2e \int^{+\infty}_{-\infty} d\tau \frac{dX^\mu_Q}{d\tau} \Theta (R)
\delta(R_\lambda R^\lambda) \eqno(10)
$$
where $R^\mu = X^\mu - X^\mu_Q$ is a null vector, 
$\tau = t_Q$ is proper time, $R = t-t_Q = | {\bf R} |$ is retardation 
condition and $\Theta (R) \delta (R_\lambda
R^\lambda) \ = \ \delta(\tau + R(\tau)-t)$. Use of the relation $\int
g(\tau)
\delta(f(\tau)-\alpha) d\tau = g(\tau) (df|d\tau)^{-1}|_{f=\alpha}$ yields$^4$ the
standard Lienard-Wiechert expressions for potentials. For uniform acceleration {\bf
g}, say along the $z$-axis, $z^2_Q-c^2 t_Q^2 = c^4/g^2 = \alpha^2$ and motion is
said to be hyperbolic ; here $df/d\tau = 1-{\bf v.n}/c = 1-c^2 t_Q \cos \theta/z_Q = \kappa$
vanishes at the boundary, so the method warrants modification. Instead, if 
$A_\mu$ is differentiated
under the integral sign, the integral expressions of the fields, yield$^{27}$ on
integration by parts the field expressions
$$
E^\Theta_R \ = \ \left[ \frac{4 e \alpha^2}{\xi^2}\right]_
{t=t_q+R} H^\Theta_\varphi = E^\Theta_\theta \ = \ 
\left[ \frac{8e \alpha^2 (t_Q+R) R \sin \theta}{\xi^3} 
\right]_{t_Q +R} \ ; \eqno(11)
$$
or in field-point co-ordinates
$$
E^\Theta_z \ = \ \frac{-4 e \alpha^2A}{\xi^3} \Theta (z+t) \ , \ 
E^\Theta_\rho \ = \ \frac{8 e\alpha^2 \rho^3}{\xi^3} 
\Theta (z+t) \eqno(12)
$$
$$
H^\Theta_\varphi = \frac{8 e \alpha^2\rho t}{\xi^3} \Theta (z+t) ; \eqno(13)
$$
where the superscript $\Theta$ refers to retardation condition and subscripts to
coordinate components ; 
$\xi^2 = (2 R \kappa z_Q)^2 = A^2 + 4 \rho^2 (z^2-t^2), A =
\rho^2 + \alpha^2 + t^2 - z^2$\,.  The fields for a charge in arbitrary 
acceleration cited in the texts$^4$, and
derivable from Lienard-Wiechert potentials were first given by Schott$^5$ and are
reducible to the above expressions. The part that is integrated out, and usually
discarded, gives here the non-vanishing contribution$^{27,8,9}$ :
$$
E^\delta_\rho \ = \ 2e \rho (\rho^2 + \alpha^2)^{-1} \delta(z+t) = -
H^\delta_\varphi\,, \eqno(14)
$$
which plays a crucial role in the considerations below.
These additional field terms with $\delta$-function were first obtained by Bondi
and Gold$^{26}$ by invoking an intricate `classical pair creation' mechanism. Already
Milner$^{23}$ in a graphical analysis of Schott fields showed that `radiation which
the solution gives is ... not from the electron..., but is to be attributed ... to
the moving boundary'. Thus computation of $\partial U^{\Theta}/\partial t +
\bigtriangledown \cdot {\bf S}^{\Theta}$ for the fields (12),(13) exhibits a charge
density on the surface $z+t=0$ corresponding to the derivative $\delta(z+t)$ of
$\Theta(z+t)$ in $\bigtriangledown . E^\Theta$ of (12), showing that there are two
charges : the original charge at $\rho = 0, z = z_Q$, and one at the boundary $z+ct=0$
; it is this latter charge which gives the outflowing radiation. The addition of
Bondi-Gold terms have the effect of cancelling this charge at the boundary and there
is no outflowing radiation.  The complete (Bondi-Gold) fields then refer$^{8,9,26,27}$
to a single charge in (perpetual) hyperbolic motion
which does not radiate$^{8,9}$. In the following absence of radiation in PUA 
is demonstrated explicitly in three different ways.

\vskip .5cm

\noindent (1) \ The Poynting vector 
represents instantaneous energy flux or power radiated. Since the $z+ct=0$
condition is relevant
only at $t \rightarrow -\infty$, both $H^\Theta_\varphi$ and
$H^\delta_{\varphi}$ vanish at $t = 0$. As {\bf H} and {\bf S} are
vectors and vanish at $t=0$, they imply, by linearity, vanishing radiation flux. This is a
covariance statement, much like the effects of magnetic field of a charge in uniform
velocity as in Trouton-Noble experiment. We conclude with Pauli$^{17}$
{\it `Hyperbolic motion thus constitutes a special case for which there is no
formation of the wave zone, nor corresponding radiation. (Radiation on
the other hand does occur when two uniform rectilinear motions are
connected by a ``portion'' of hyperbolic motion)'}. Strangely, however,
this argument of Pauli has not found much favour with proponents of
radiation hypothesis. In the following, this statement is explicitly
verified in toto.

\noindent (2) \ Let the volume element $dV=d\rho \wedge \rho d \varphi
\wedge d z = d R \wedge d\Sigma_R = \kappa R^2 dR \wedge d\Omega, d\Omega =
\sin \theta d\theta \wedge d\varphi$. Then for the fields (11)
$$
\lim_{\stackrel{t=R\rightarrow \infty}{t_Q =0}} \frac{dW^\Theta}{dt} \
= \ \lim_{\stackrel{R \rightarrow \infty}{t={\rm constant}}}
\frac{-dW^\Theta}{dt_Q} \ = \ \frac{2e^2}{3\alpha^2} \ = \ \lim_{R
\rightarrow \infty} \int {\bf S}^\Theta \cdot d\Sigma_R = P^\Theta
\eqno(15)
$$
Similarly for the fields (12) and (13)
$$
\lim_{t\rightarrow \infty} \frac{dW^\Theta}{dt} \ = \
\frac{2e^2}{3\alpha^2}\,. \eqno(16)
$$
For the complete fields $W = W^\Theta + W^{\Theta \delta} + W^\delta$.
While $W^\delta$ is not too well-defined, it is independent of time, and
$$
W^{\Theta \delta} \ = \ \frac{-2}{3} \frac{e^2}{\alpha^2} t \ , \ \int
dx \Theta (x) \delta (x) {\stackrel{def}{=}} \frac{1}{2}\,, \eqno(17)
$$
so that $dW/dt=0$ for the complete fields, showing that there is no
outgoing radiation loss$^{8,9}$.

\vskip .5cm

Likewise in the total power radiated for the complete fields $P = P^\Theta +
P^{\Theta \delta} + P^\delta, P^\delta$ is ill-defined, but is taken
care of in the identity (6), and ignored here. The $z$ and $\rho$
components of {\bf S}$^{\Theta \delta}$ are $\frac{c}{4\pi}
E^\delta_\rho (H^\Theta_\varphi - E^\Theta_\rho)$ and
$\frac{-c}{4\pi} E^\delta_\rho E^\Theta_z$, so
$$
P^{\Theta \delta} \ = \ 8ce^2 \alpha^2 \int \frac{\rho^3
d\rho}{\rho^2 + \alpha^2} \frac{t-z}{\xi^3} \Theta \delta \left|_z
-4ce^2 \alpha^2 \int \frac{\rho^2 A \Theta \delta}{(\rho^2 +
\alpha^2) \xi^3} dz \right|_{\rho \rightarrow \infty} \,, \eqno(18)
$$
where the second term vanishes as $\rho \rightarrow \infty$ ; the first
term, containing the bare (un-integrated) $\Theta \delta$ terms are to
be evaluated on the {\it lower} surface $z+ct=0$ ; we interpret this as
$$
- \frac{t-z}{\xi^3} \Theta \delta \left|_{z+ct}
\stackrel{\rightarrow}{=} \frac{2t}{(\rho^2 + \alpha^2)^3} 
\frac{-1}{2t} \ = \ 
\frac{-1}{(\rho^2 + \alpha^2)^3} \,, \right. \eqno(19)
$$
to obtain, after integration over $\rho$, on restoring $\alpha^2 = c^4/g^2$
$$
P^{\Theta \delta } \stackrel{\rightarrow}{=} \frac{-2}{3} \frac{e^2
g^2}{c^3} ; \eqno(20)
$$
this complements $P^\Theta$ in (15) to give $P = P^\Theta+
P^{\Theta \delta} =0$ and vanishing power radiated.

\vskip .5cm

\noindent (3) \ Consider now the power spectrum. As a first
step take the Fourier transform of (18)
$$ \begin{array}{lcl}
\hat{P}^{\Theta \delta}(\omega) & = & 8ce^2 \alpha^2 \left\{ \int
\frac{\rho^3 d\rho}{\rho^2+\alpha^2} \int dt \frac{t-z}{\xi^3}
\Theta \delta e^{i\omega t} \left\vert_{z+ct=0} 
- \frac{\rho^2}{\rho^2+\alpha^2} \int \frac{A
\Theta \delta}{\xi^3} dz dt e^{i\omega t} \right\vert_{\rho
\rightarrow \infty}\right\} \\ 
& = & \left. \frac{-2}{3} \frac{e^2 {\bf g}^2}{c^3} \int dt (z-t) \theta \delta 
e^{i\omega t} \right\vert_{z+ct=0} - 2\pi
\delta(\omega) \lim_{\rho \rightarrow \infty} \frac{\rho^2}{(\rho^2
+ \alpha^2)^3} \end{array} \,, \eqno(21) 
$$
which shows that it is non vanishing only at $\omega =0$, in 
accord with the above and earlier preliminary considerations.

\vskip .5cm

Complementary to the above analysis is the spectral analysis
of $P^\Theta$. Notice that $dP^\Theta = {\bf S}^\Theta .
d\Sigma_R = \frac{c}{4\pi} (RE^\Theta_\theta)^2 d \Omega$, so
angular distribution of power is $dP/d\Omega =
c(RE^\Theta_\theta)^2/4\pi$, and, by  Plancheral theorem, its'
spectral resolution is
$$
\frac{d\hat{P}(\omega)}{d\Omega} \ = \ \frac{c}{4\pi} \cdot
\frac{1}{2\pi} \left\vert \int RE^\Theta_\theta e^{i\omega t}
dt\right\vert^2\, ; \eqno(22)
$$
$t=t_Q+R$ and $dt/dt_Q=\kappa$ ; in $RE^\Theta_\theta =
\frac{e\alpha^2}{\kappa^3 z^3_Q} (1+t_Q/R)\,,$ omitting the near field
term $t_Q/R$,
$$ \begin{array}{lcl}
\frac{d\hat{P}^\Theta(\omega)}{d\Omega} & = & \frac{c}{4\pi} \cdot
\frac{1}{2\pi} \left\vert \int \frac{e\alpha^2}{\kappa^2 z_Q^3}
e^{i\omega(t_Q +R)} dt_Q \right\vert^2  \\
& = & \frac{c}{2(2\pi)^2} \left\vert \left[ \frac{ev \sin\theta}{\kappa}
e^{i\omega t}\right] - ie \int \omega v \sin \theta e^{iwt} dt_Q
\right\vert^2\,.\end{array} \eqno(23) 
$$

Inserting $t = t_Q+r-{\bf n}.{\bf x}_Q, {\bf n} = {\bf x}/r$ for large $r = |{\bf
x}|$, and dropping the integrated term, in non-relativistic limit 
$ {\bf x}_Q = \frac{1}{2} {\bf g} t_Q^2, {\bf v} = {\bf g} t_Q$, yields 
$$
\frac{d\hat{P}(\omega)}{d\Omega} = \frac{e^2\omega^2({\bf g} \times {\bf n})^2}{4\pi}
\int^{+\infty}_{-\infty} \left[T^2 \delta (\omega \kappa) + \frac{1}{4}
\delta '' (\omega \kappa)\right] dT \eqno(24)
$$
where the substitutions $t_{Q1} = T - \frac{1}{2} T',
t_{Q2} = T + \frac{1}{2} T'$, 
and the definition of the Dirac delta function.
are used in the double integral (24), and $\kappa
= 1-{\bf n}.{\bf v}(T)/c = 1-{\bf n}.{\bf g}T/c.$ 
Clearly the integrand in
(24), 

$$\frac{d\hat{P}^\Theta (\omega,T)}{d\Omega} = 0 \quad {\rm for \ all} \quad 
\omega > 0 \ {\rm as} \ \kappa \ne 0 ; \eqno(25)
$$
i.e. power is non-zero only for $\omega=0$ which
corresponds to the non-radiation case. Note that though (21) and
(24) have the same meaning, they are not formally identical, as (21)
is a plain Fourier transform while (25) utilizes Plancheral
theorem.  For the relativistic analogue of (23), dropping the integrated term in
(23), the change of variable $vdt_Q = dz_Q$, with $z_Q=(\alpha^2 + c^2 t^2_Q)^{1/2}$
and $t \approx r+t_Q - z_Q \cos \theta$ yields 
$$ \begin{array}{lcl}
\frac{d\hat{p}^\theta (\omega)}{d\Omega} & = & \frac{c}{2} \left( \frac{e \omega
Sn\theta}{\pi}\right)^2 \left| \int^\omega_\alpha e^{-i\omega c^{-1} \cos \theta
z_Q} \sin (\omega^{-1} c\sqrt{z_Q^2 - \alpha^2} dz_Q \right|^2 \\
& = & \frac{ce^2}{2\pi^2} (\alpha w K_1 \left( \frac{\alpha \omega \sin
\theta}{c}\right))^2 \,, \end{array} \eqno(26)
$$
where $K_1$ is the modified Bessel function (of the second kind) ; for $0 \le x < 2$
$x K_1(x)$ falls off steeply from 1 at $x=0$ to  $\sim 0.3$ at $x=1$, somewhat like eqn.(9)
near $\omega = 0$, and tends to zero as $xe^{-x}$, for large $x$. Comparing with the
case of ``transverse contraction' of the field of a relativistic charge$^{28}$ in
uniform velocity, the corresponding effective `time of passage' is of the order $c
\sin \theta/2\pi g$, with
$$
\frac{d\hat{p}^\theta (\omega)}{d\Omega}  = 0 \qquad {\rm for} \quad \omega >
\frac{c}{\alpha \sin \theta} = \frac{g}{c \sin \theta} \eqno(27)
$$
This is so if one defines, as in eqn.(9), $\omega_o = 2\pi/T$. Since maximum
velocity may not exceed $c > gT$, there is a bound $\omega > \frac{2\pi g}{c}$, with
$\sin \theta \sim (2\pi)^{-1}$ ; this is reasonable as angle of `maximum intensity'
for relativistic velocities is $\sim \frac{1}{2} \sqrt{1-v^2/c^2} \sim$ the ratio of
rest energy to its total energy.

Both (20) and (21), marred by singular analogies, are meant only as symptomatic of
non-relativistic results (9) and (25). The relativistic treatment above,
eqn.(26,27), already shows the effects of relativistic attenuation and infinitely
long time interval for acceleration. Indeed, a more exact treatment, including the
near field effects, using Plancheral theorem, for the complete fields should bring
out the broad features encountered above : viz. Larmor formula represents radiation
loss {\it only if} there is change in acceleration, shown by a characteristic peak
in the Spectral resolution ; For the complete fields representing Hyperbolic motion
there is no energy loss whatsoever.  

\vskip .5cm

It is of interest to compare the case of uniform velocity. Here, there is
the notion of kinetic energy, which, however, quantitatively depends on
the inertial frame chosen unless there is change in velocity. With
change in velocity kinetic energy has a nontrivial quantitative and
dynamical meaning. Same is the situation for the Larmor formula which has a
quantitative meaning only when there is change in acceleration. The analogy
goes deeper : Lorentz velocity boosts together with spatial euclidean
motions and time translations are transitive on homogenous space-time
metric $c^2 dt^2 - d{\bf x}^2$ of SR, just as acceleration hikes
together with spatial euclidean motions and space-time dilations leave
unchanged$^{12}$ the homogeneous space-time metric $(c^2 dt^2 - d{\bf
x}^2)/t^2$. Thus we see that uniform acceleration and equivalence
principle take us right into the heart of general relativity giving the
metric of space-time basic to the notion of a cosmological constant. 

\vskip .5cm

\noindent {\bf To summarise} \ (1) The paradox arises only if one fails to
distinguish between the case of uniform acceleration for a finite period from
perpetual uniform acceleration. Radiation does occur in the first case, but not in
the second case. This is like the clock paradox which arises if one does not
distinguish between perpetual and non-perpetual inertial frames. (2) There is no
need to postulate Acceleration$^{13}$ or Schott$^{14}$ energy or question$^{15}$ 
the validity of Equivalence princple or of Conformal symmetry argument and attempt 
to$^{14,15,17}$ replace these
by certian flat space transformations with associated hypothesis of event
horizons to render radiation unobserable$^{16}$.

\newpage

\noindent {\bf REFERENCES}

\vskip .5cm

\begin{enumerate}
\item J.Larmor, {\it On Dynamical Theory of the Electric and Luminiferous Medium},
Part III {\it Relations with Material Media}, Phil. Trans. Roy. Soc. Lond. 
{\bf 190A} (1897) 205-230.
\item J.Larmor, {\it Aether and Matter} (1900) Cambridge U.P.
\item J.Larmor, {\it On the Theory of the Magnetic Influence on Spectra ; And On the
Radiation from Moving Ions}, Phil. Mag. {\bf 44}(S.5) (1897) 503-512.
\item J.D.Jackson, {\it Classical Electrodynamics} (1962) Wiley, N.Y.
\item G.A.Schott, {\it Electromagnetic Radiation} (1912) Cambridge, U.P.
\item K.H.Mariwalla, {\it Vectors, Tensors and Relativity}, Matscience Report No.84
(Ed. A.Ramakrishnan, I.M.Sc., Madras 1975).
\item R.H.Dicke, {\it Experimental Relativity} in {\it Relativity, Groups and Topology} 
(Eds. De Witt and De Witt, N.Y. 1963) p.168. 
\item K.H.Mariwalla, J.J.Walker, {\it Radiation Emitted by a Uniformly 
Accelerated Charge}, Bull. Am. Phys. Soc. {\bf 10}, 259 (1965).
\item K.H.Mariwalla, R.Vasudevan, {\it Does a Uniformly Accelerated Charge Radiate?},
Lett. Nuo. Cim. {\bf 1}, 225-228 (1971).
\item E.L.Hill, {\it On Accelerated Coordinate Systems in Classical Relativistic
Mechanics}, Phys. Rev. {\bf 67}, 358-363 (1945).
\item K.H.Mariwalla, {\it Coordinate Transformations that Form Groups In The Large}, 
in {\it Boulder Lectures} Vol. VIII, 177-192 {\it (de Sitter and Conformal 
Groups},  Eds. A.O.Barut, W.E.Brittin, Colorado  Associated UP. 1970).
\item K.H.Mariwalla, {\it Dynamical Symmetries In Mechanics}, Physics Reports 
{\bf 20C}, 287-362 (1975).
\item G.A.Schott, {\it On the Motion of the Lorentz Electron}, Phil. Mag. (S.6)
{\bf 29}, 49-62 (1915).
\item T.Fulton, F.Rohrlich, {\it Classical Radiation from Uniformly Accelerated
Charge}, Ann. Phys. (N.Y.) {\bf 9}, 499-517 (1960). 
\item F.Rohrlich, {\it The Principle of Equivalence}, Ann. Phys. {\bf 22}, 169-191
(1963).
\item Rudolf Peierls, {\it Radiation in Hyperbolic Motion} (Sec.8.1) in 
{\it Surprises in Theoretical Physics}, (Princeton U.P., Princeton, 1979).
\item W.Pauli, {\it Theory of Relativity} (Pergamon Press, London, 1958). Original German (1918).
\item K.H.Mariwalla, R.Vasudevan, {\it Uniform Acceleration in Special Relativity},
in Particle Interactions and Astrophysics, 225-228 Matscience Report {\bf 108}, 
(Conference Proceedings, Ed.  A.Ramakrishnan, I.M.Sc., Madras, 1981) 
\item R.A.Mould, {\it Internal Absorption Properties of Accelerating Detectors} 
Ann. Phys. (NY) {\bf 27}, 1-12 (1964).
\item R.O.Grigo'rev, {\it Quantization of Systems in Incomplete Spaces and the
Problem of Particle Production by a Gravitational Field}, Theo. Math. Phys. 
{\bf 89}, 1348-1353 (1992) ( Teo. i Mat. Fizika {\bf 89} 473-480 (1991), Russian).
\item F.Hinterleitner, {\it Inertial and Accelerated Particle Detectors with Back
Reaction in Flat Space-Time} Ann. Phys. (NY) {\bf 226}, 165-204 (1993).
\item M.T.Jaekel, S.Reynaud, {\it Vacuum Fluctuation, Accelerated Motion and
Conformal Frames}, Quant. Semiclass. Opt. {\bf 7}, 499-508 (1995).
\item S.R.Milner, {\it Does an Accelerated Electron necessarily Radiate Energy on
the Classical Theory}, Phil. Mag. (S.6){\bf 41}, 405-419 (1921).
\item M.v.Laue, {\it Relativit\"{a}tstheorie}, 3rd. ed. vol.1, Viewet, Braunschwieg
(1919).
\item H.Bondi, {\it The Field of a Uniformly Accelerated Charge with Special
Reference to the Problem of Gravitational Acceelration}, Helvetica Physica Acta, 
Suppl. IV (1956) p.98 ({\it Jubilee of Relativity}, Ed. A. Mercier, M.Kervaire).
\item H.Bondi, T.Gold, {\it The Field of a Uniformly Accelerated Charge with Special
Reference to the Problem of Gravitational Acceelration}, Proc.  Roy. Soc. 
(Lond.) {\bf A229},  416-424(1955).
\item J.J.Walker, {\it Electromagnetic Field of a Charge in Hyperbolic Motion},
Bull. Am. Phys. Soc. {\bf 10}, 259 (1965).
\item W.K.H.Panofsky, M.Phillips, {\it Classical Electricity and Magnetism},
Sec.18.5, p.295 (Addison-Wesley, Reading, Mass., 1955-56).
\end{enumerate}

\vskip .5cm

\noindent {\bf Acknowledgements } \ KHM acknowledges
collaboration over the years with E C Lerner$^{8}$, J J Walker$^{8,27}$, 
R Vasudevan$^{9,18}$ and C J
Eliezer. Thanks are due to Rahul Sinha for a copy of Coleman's preprint, and
to H S Sharatchandra and R Anishetty for conversations. NDH is obliged to Kim Milton
for making available a section of the manuscript of J. Schwinger, L.L.
De Raad Jr, K.A. Milton and
W.Y. Tsai.
\vskip .5cm
Correspondence and request for materials to the authors, Institute of
Mathematical Sciences, C.I.T campus, CHENNAI 600 113, INDIA.\\
e-mail: mari@imsc.ernet.in,dass@imsc.ernet.in
\end{document}